\documentclass[10pt]{iopart}

\usepackage{epsfig}
\usepackage{graphicx}
\usepackage{amssymb}
\usepackage{iopams}
\usepackage{latexsym}
\usepackage{fontenc}
\usepackage{geometry}
\usepackage{setspace}
\usepackage{cite}
\usepackage{soul}

\usepackage[hyperindex=true,
pdftitle={Modeling of the electromagnetic fields of elementary particles in Einstenian gravity},
colorlinks=true,
pagebackref=false,
citecolor=blue,
plainpages=false,
pdfpagelabels,
breaklinks=true,
linkcolor=blue,
urlcolor=blue,
]{hyperref}
\usepackage{microtype}

\newcommand{\I}{{\rm i}}

\newcommand{\be}{\begin{equation}}
\newcommand{\ee}{\end{equation}}
\newcommand{\bea}{\begin{eqnarray}}
\newcommand{\eea}{\end{eqnarray}}
\newcommand{\nn}{\nonumber}

\def\Xint#1{\mathchoice
{\XXint\displaystyle\textstyle{#1}}%
{\XXint\textstyle\scriptstyle{#1}}%
{\XXint\scriptstyle\scriptscriptstyle{#1}}%
{\XXint\scriptscriptstyle\scriptscriptstyle{#1}}%
\!\int}
\def\XXint#1#2#3{{\setbox0=\hbox{$#1{#2#3}{\int}$}
\vcenter{\hbox{$#2#3$}}\kern-.5\wd0}}

\def\dashint{\Xint-}
\def\lbar{\lambda{\hspace{-0.17cm}^{_{^-}}}}

\def\nn{\nonumber}

\begin{document}

\title[]{\bf The influence of the Lande $g$-factor in the classical general relativistic description
of atomic and subatomic systems}
\author{Leonardo A Pach\'{o}n$^{1,2}$ and F L Dubeibe$^{1,3}$}
\address{$^{1}$Departamento de F\'isica, Universidad Nacional de Colombia, Bogot\'a, Colombia\\
$^{2}$ Department of Chemistry, University of Toronto, Toronto, Ontario, Canada M5S 3H6 \\
$^{3}$Facultad de Ciencias Humanas y de la Educaci\'on, Escuela de Pedagog\'ia, \\
Universidad de los Llanos, Villavicencio, Colombia}

\eads{\mailto{lapachonc@unal.edu.co}, \mailto{fdubeibe@gmail.com}}
\begin{abstract}
We study the electromagnetic and gravitational fields of the proton and electron in terms of the
Einstenian gravity via the introduction of an arbitrary Lande $g$-factor in the Kerr-Newman solution.
We show that at length scales of the order of the reduced Compton wavelength, corrections from different
values of the $g$-factor are not negligible and discuss the presence of general relativistic effects
in highly ionized heavy atoms. On the other hand, since at the Compton-wavelength scale the gravitational
field becomes spin dominated rather than mass dominated, we also point out the necessity of including
angular momentum as a source of corrections to Newtonian gravity in the quantum description of gravity
at this scale.
\end{abstract}


\section{Introduction}
Since the construction of a well-defined and well-established quantum theory for gravity remains
still as an open problem, questionings about quantum mechanical corrections to Newton's gravitational
and Coulomb's electrostatic potentials at subatomic scales are completely legitimate\footnote{Abusing
terminology, hereafter we will refer to the Coulomb and Newton potentials as classical potentials}.
In this respect, corrections in the framework of causal perturbation theory \cite{EG73} including
one-loop contributions to graviton self-energy \cite{ HPS00, EL&95, gri01, bje02, BDH03, but06, fal08},
suggest that quantum corrections could be relevant at length scales of the order of the reduced
Compton wavelength $\lbar_{\rm C} = \hbar/mc$ for the electrostatic potential \cite{HPS00} and at
the order of the Planck length $l_{\rm P} = \sqrt{G \hbar/c^3}$ for the gravitational potential
\cite{EL&95,gri01}.

These questionings are also valid from a general relativistic point of view, \textit{i.e.}, could
general relativistic effects be relevant in the description of atomic and subatomic systems? This
conundrum has been addressed by Martin and Pritchett in a seminal work about the role of the
magnetic dipole in the gravitational field of the electron \cite{MP68} and subsequently by several
authors \cite{isr70,NW74,bur74,lop84,MM93,bur03,AP04,bur04,bur05,ros06}. Additionally, Carter 
\cite{car68} and Rosquist \cite{ros06} pointed out that general relativistic effects become relevant 
at the reduced Compton wavelength scale for Coulomb's electrostatic potential while Newman {\it{et.al}}  
\cite{NC&65} found that these effects become important at the Planck length scale for Newton's gravitational
potential. However, these estimations are based on Kerr-Newman's solution \cite{NC&65},
which has gyromagnetic ratio $g_{\rm KN} = 2$ \cite{car68}; albeit, even for the electron, the real
value of the gyromagnetic ratio differs from 2, $g_{\rm e}=$2.0023193043768(86) \cite{MTN08}, and
for the proton the real value is {\it ca.} three times $g_{\rm KN}$, $g_{\rm p}=$5.585694713(46)
\cite{MTN08}. For this reason, a more general model which allows to study corrections from $g\ne2$
is desirable.

In order to get an idea of how large the effects of an arbitrary $g$--factor may be for an extended
spinning particle, we use as a toy model the simplest generalization of the Kerr-Newman solution
derived by Manko \cite{mak93}. We introduce the gyromagnetic ratio as a free parameter in
\cite{mak93} and after an asymptotic expansion of the full solution, we find that corrections to
the electric field of the electron presented by Rosquist \cite{ros06} using the Kerr-Newman solution,
remain practically unaffected by using the real value of the $g$--factor. However, this is
not longer true for the proton fields which certainly differ from the Kerr-Newman based description.

At this point, two natural questions arise: \textit{i}) If both theories, quantum mechanics and general
relativity, predict corrections at the same length scale, which one is stronger than the other? and
\textit{ii}) could be possible to detect these corrections, e.g., in atomic or subatomic systems?
Concerning the first one: We compare our results with those obtained in Refs. \cite{bje02,BDH03,fal08}
for the Coulomb and Newton potentials at length scales of the reduced Compton wavelength of the proton
and electron, and find that corrections from general relativity are stronger than the ones given by the
quantum approach for gravity. Concerning the second one: We introduce our corrections in Bohr's
description of hydrogen-like atoms and since atomic radii are far away from the region where deviations
from the classical treatment are expected, we just find that corrections in this case are below the
uncertainty of the calculations\footnote{Taking into account the uncertainties for the fundamental constants}.

We have organized the paper as follows. In section \ref{sec2} some remarks about the solution by Manko
\cite{mak93} are presented and the $g$-factor is introduced. In section \ref{sec3}, we present the
asymptotic expansion of the electromagnetic potentials of the solution by Manko. In section \ref{sec4},
for the case of the electron and proton, deviations of the electromagnetic and gravitational fields
using the real value of the $g$-factor are analyzed, the possibility of correction in atomic systems
is also discussed; we close this section with the comparison of our results with those obtained from
causal perturbation theory \cite{BDH03,fal08}. Finally, we present some concluding remarks in section
\ref{sec5}.

\section{General relativistic description of a rotating charged magnetic dipole}\label{sec2}
The simplest metric describing the geometry of spacetime around a stationary and axisymmetric source
is given by Papapetrou's line-element
\begin{eqnarray}
    \label{metric}
    ds^2= -F(dt-\omega d\phi)^2+F^{-1}\left[e^{2\gamma}(d\rho^2+dz^2)+
    \rho^2 d\phi^2 \right]\, ,
\end{eqnarray}
where the metric functions $F$, $\gamma$ and $\omega$ depend only on the Weyl-Papapetrou quasi-cylindrical
coordinates $\rho$ and $z$. The associated Einstein-Maxwell field equations defining the metric
functions in (\ref{metric}) can be reformulated in terms of the Ernst complex potentials
$\mathcal{E}(\rho,z)$  and $\Phi(\rho,z)$ (see \cite{ern68} for details). With the aid of Sibgatullin's
integral method \cite{MS93,sib91}, the Ernst potentials can be
calculated from specified axis data $e(z):={\cal E}(z,\rho=0)$ and $f(z):=\Phi(z,\rho=0)$, by the
integrals
$ {\cal E}(z,\rho)=\frac{1}{\pi}\int_{-1}^1 \frac{e(\xi)\mu(\sigma)d\sigma}{\sqrt{1-\sigma^2}},$
and
$\Phi(z,\rho)=\frac{1}{\pi}\int_{-1}^1 \frac{f(\xi)\mu(\sigma)d\sigma}{\sqrt{1-\sigma^2}}\, .$
The unknown function $\mu(\sigma)$ must satisfy the singular integral equation
$\dashint_{-1}^{1}\frac{\mu(\sigma)[e(\xi)+\tilde{e}(\eta) +
2f(\xi) \tilde{f}(\eta)]d\sigma}{(\sigma-\tau)\sqrt{1-\sigma^2}}=0\, ,
$
and the normalization condition
$
\int_{-1}^1\frac{\mu(\sigma)d\sigma}{\sqrt{1-\sigma^2}}=\pi,
$
where $\xi=z+i\rho\sigma$, $\eta=z+i\rho\tau$ and $\sigma, \tau\in[-1,1]$,
$\tilde e(\eta):=[e(\eta^*)]^*$, $\tilde{f}(\eta):=[f(\eta^*)]^*$ and the star stands for complex
conjugation.

For a rotating charged magnetic dipole, Ernst's potentials on the symmetry axis can be taken as
\begin{eqnarray}
\label{Ernstaxis}e(z)=\frac{z-m-ia}{z+m-ia}\, , \qquad
f(z)=\frac{qz+ib}{z(z+m-ia)}\, .
\end{eqnarray}
For this case, the Ernst potentials and the corresponding metric functions were derived by Manko in
\cite{mak93}. However, the original paper has some minor typos and we consider appropriate to write
the full expressions in this paper in order to enable the solution for further studies\footnote{We
thank Prof. Manko for providing a rectified version of the metric function $\omega$.}. The Ernst
potentials of the solution are given by
\begin{eqnarray}
  \label{ErnstPotentials}
\hspace{-2.5cm}    {\cal E}&=&\frac{A-B}{A+B},\quad
    \Phi=-\frac{C}{A+B},
\end{eqnarray}
with
\begin{eqnarray}
\hspace{-2.5cm}    A&=&\kappa_-^2[(\kappa_+^2-a q-b)(R_- r_- + R_+ r_+)
    + \I \kappa_+ (a + q)(R_- r_- - R_+ r_+)]
\nn \\
\hspace{-2.5cm}    &+&\kappa_+^2[(\kappa_-^2 + a q + b)(R_- r_+ + R_+ r_-)
    + \I \kappa_- (a - q)(R_- r_+ - R_+ r_-)] -  4 b(a q + b)(R_- R_+ + r_- R_+),
\nn \\
\hspace{-2.5cm}    B&=&m\kappa_+ \kappa_- \{(m^2 - a^2 - q^2)(r_- + r_+ - R_- -
    R_+) + \kappa_+ \kappa_- (r_- + r_+ + R_- + R_+)
\nn \\
\hspace{-2.5cm}    &+& \I q[(\kappa_+ + \kappa_-)(r_- - r_+) + (\kappa_+ - \kappa_-)(R_+ - R_+) ] \},
\nn \\
\hspace{-2.5cm}    C&=& \kappa_+ \kappa_-\{[q(m^2 - a^2 - q^2)-2 a b](R_- + R_+ - r_- -
    r_+) - q \kappa_+ \kappa_-(R_- + R_+ + r_+ - r_-)
\nn \\
\hspace{-2.5cm}    &+& \I[\kappa_+(q^2 + b)(R_- - R_+ + r_+ - r_-)
    + \kappa_-(q^2 - b)(R_+ - R_-  + r_+ - r_-) ] \},
\nn \\
\hspace{-2.5cm}    R_\pm&=&\sqrt{\rho^2+[z \pm {\scriptstyle \frac{1}{2}}(\kappa_+ + \kappa_-)]},
\quad
    r_\pm = \sqrt{\rho^2+[z \pm {\scriptstyle \frac{1}{2}}(\kappa_+ - \kappa_-)]}, \quad
    \kappa\pm=\sqrt{m^2 - a^2 - q^2 \pm 2 b}. \label{solution}
\end{eqnarray}
The metric functions $F,\gamma$, and $\omega$ entering into (\ref{metric})
are given by
\begin{eqnarray}
    F&=&\frac{A {\bar A} - B {\bar B} + C {\bar C}}{(A + B)({\bar A}+
    {\bar B})}\, ,
\quad
    {\rm e}^{2 \gamma}=\frac{A {\bar A} - B {\bar B} + C {\bar C}}{16[(m^2 - a^2 - q^2)^2
    - 4b^2]^2 R_- R_+ r_- r_+ }\, ,
\end{eqnarray}
\begin{eqnarray}
    \omega&=&\frac{{\rm Re}[C D + \I[(E - q C)({\bar A} + {\bar B}) + m G(A + B)]]}
    {A {\bar A} - B {\bar B} + C {\bar C}}\, ,
\end{eqnarray}
with
\begin{eqnarray}
\hspace{-2.5cm}    D&=& m\{\kappa_-^2[(b - q^2)(R_- r_- + R_+ r_+) -\I q \kappa_+ (R_- r_- - R_+ r_+)]
+   \kappa_+^2[(b + q^2)(R_- r_+ + R_+ r_-)
\nn \\
\hspace{-2.5cm}    &-&\I q \kappa_- (R_- r_+ - R_+ r_-)]-2 b[(m^2 - a^2 + q^2)(R_- R_+ + r_+ r_+) +
\kappa_- \kappa_+(R_- R_+ - r_- r_+) + 2 \kappa_-^2 \kappa_+^2]\}
\nn \\
\hspace{-2.5cm}    &+&\kappa_- \kappa_+\{[b(3 m^2 + a^2 + q^2 - 2 \I a z) - q(m^2 - a^2 - q^2)(a - \I
    z)](r_- - R_- + r_+ - R_+)
\nn \\
\hspace{-2.5cm}    &-&(b + a q -\i a z)[\kappa_- \kappa_+(R_- + R_+ + r_- + r_+) -
\I q (\kappa_+ + \kappa_-)(r_- - r_+)
\nn \\
\hspace{-2.5cm}     &-& \I q(\kappa_+ - \kappa_-)(R_+ - R_-)] + b z[(\kappa_+ - \kappa_-)(r_- - r_+) +
     (\kappa_+ + \kappa_-)(R_+ - R_-)] \},
\nn \\
\hspace{-2.5cm}    E&=&\kappa_-^2\{[\kappa_+^2(m + \I q) - (b + a q)(m - \I a)](R_- r_- + R_+ r_+)
    +\kappa_+[m^2 - q^2 + b +\I m(a + q)](R_- r_- - R_+ r_+) \}
\nn \\
\hspace{-2.5cm}    &+&\kappa_+^2\{[\kappa_-^2(m - \I q) + (b + a q)(m - \I a)](R_- r_+ + R_+ r_-)
    +\kappa_-[m^2 - q^2 - b +\I m(a - q)](R_- r_+ - R_+ r_-) \}
\nn \\
\hspace{-2.5cm}    &-&4 b(m- \I a)(b + a q)(R_- R_+ + r_- r_+)
\nn \\
\hspace{-2.5cm}    &+&\kappa_- \kappa_+\{[(m^2 - a^2 + q^2 - \I m a)(R_- + R_+ + r_- + r_+) - 2 a q
    b](R_- - r_- + R_+ - r_+)
\nn \\
\hspace{-2.5cm}    &-&\kappa_- \kappa_+(m z -m^2 + q^2 - \I m a)(R_- + R_+ + r_- + r_+)
\nn \\
\hspace{-2.5cm}    &+&\I q (m z -m^2 + q^2 - \I m a)[(\kappa_+ - \kappa_-)(R_- - R_+)
+ (\kappa_+ + \kappa_-)(r_+ - r_-)]
\nn \\
\hspace{-2.5cm}    &+&\I q  b[(\kappa_+ + \kappa_-)(R_- - R_+) + (\kappa_+ - \kappa_-)(r_+ -
    r_-)]\},
\nn \\
\hspace{-2.5cm}    G&=&\kappa_- \kappa_+\{[(m^2 - a^2 - q^2)(z + \I a) - 2 \I q b](r_- - R_- + r_+ - R_+)
+  \kappa_-\kappa_+ (z + \I a)(R_- + R_+ + r_- + r_+)
\nn \\
\hspace{-2.5cm}    &+&(b + a q -\I q z)[(\kappa_+ + \kappa_-)(r_- - r_+) + (\kappa_+ - \kappa_-)(R_+ - R_-)]\}
\nn \\
\hspace{-2.5cm}    &+&m\{\kappa_+^2[(\kappa_- - \I q)R_+ r_- - (\kappa_- + \I q)R_- r_+] +
    \kappa_-^2[(\kappa_+ + \I q)R_+ r_+ - (\kappa_+ - \I q)R_- r_-]
\nn \\
\hspace{-2.5cm}
    &+& 4 \I q b(R_+ R_- + r_+ r_-)\}.
\end{eqnarray}
Note that $F,\omega$, and $\gamma$ satisfy $\omega \rightarrow 0$, $F \rightarrow 1$ and
$\gamma \rightarrow 0$ at infinity.

\section{Asymptotic expansion of gravitational and the electromagnetic potentials}\label{sec3}
In order to understand the physical meaning of the four arbitrary parameters $m$, $a$, $q$ and $b$ in
(\ref{Ernstaxis}) and to calculate the approximate potentials  it is helpful to change the potentials
$\mathcal{E}$ and $\Phi$ to the potentials $\xi$ and $\varsigma$ --which  are related to the gravitational
and electrostatic potentials in a very direct way (see Eqs. (\ref{xi}) and (\ref{varsigma}))--
throughout the following definitions:
\begin{eqnarray}
    \label{PotTrans}
    {\cal E}=\frac{1-\xi}{1+\xi}, \quad \Phi=\frac{\varsigma}{1+\xi}.
\end{eqnarray}
These potentials satisfy \cite{SA04}
\begin{eqnarray}
\label{eq:GenerLaplaceXI}
(\xi \xi^{*} - \varsigma \varsigma^{*} -1)\nabla^2\xi =
2( \xi^{*} \nabla\xi -  \varsigma^{*} \nabla\varsigma)\cdot \nabla\xi,
\\
\label{eq:GenerLaplaceSIGMA}
(\xi \xi^{*} - \varsigma \varsigma^{*} -1)\nabla^2\varsigma =
2( \xi^{*} \nabla\xi -  \varsigma^{*} \nabla\varsigma)\cdot \nabla\varsigma.
\end{eqnarray}
Since this set of equations denotes an alternative representation of the Einstein-Maxwell field
equations, they could be understood as the generalization of Laplace's equation for (\ref{metric}).

In order to measure the moments of an asymptotically flat spacetime, according to the Geroch-Hansen
procedure \cite{ger70,han74}, we map the initial 3-metric to a conformal one $h_{ij} \rightarrow
\tilde{h_{ij}} = \Omega^{2}h_{ij}$. The conformal factor $\Omega$ should satisfy the following
conditions:
$\Omega|_{\Lambda} = \tilde{D_{i}}\Omega|_{\Lambda} = 0$ and
$\tilde{D_{i}}\tilde{D_{j}}\Omega|_{\Lambda}= 2h_{ij}|_{\Lambda}$, where $\Lambda$ is the point added
to the initial manifold that represents infinity. $\Omega$ transforms the potentials $\xi$ and
$\varsigma$ into $\tilde{\xi} = \Omega^{-1/2}\xi$ and
$\tilde{\varsigma} = \Omega^{-1/2}\varsigma$, respectively. The conformal factor is given by
$\Omega = \bar{r}^{2} = \bar{\rho}^{2} + \bar{z}^{2}$, and the transformation relation between the
barred and unbarred variables is given as
\begin{eqnarray}
\bar{\rho}=\frac{\rho}{\rho^{2}+z^{2}},\quad \bar{z}=\frac{z}{\rho^{2}+z^{2}}, \quad \bar{\phi}=\phi,
\end{eqnarray}
which brings infinity at the origin of the axes $(\bar{\rho},\bar{z}) = (0, 0)$. The potentials
$\tilde{\xi}$ and $\tilde{\varsigma}$ can be written in a power series expansion of  $\bar{\rho}$ and
$\bar{z}$ as
\begin{eqnarray}\label{xi-q}
\tilde{\xi} = \sum_{i,j=0}^{\infty} a_{ij}\bar{\rho}^{i}\bar{z}^{j},\quad
\tilde{\varsigma} = \sum_{i,j=0}^{\infty}  b_{ij}\bar{\rho}^{i}\bar{z}^{j}.
\end{eqnarray}
Due to the analyticity of the potentials at the axis of symmetry, $a_{ij}$ and $b_{ij}$ must vanish
when $i$ is odd \cite{SA04,PS06}. The coefficients in the above power series can be calculated by the
relation \cite{SA04}
\begin{eqnarray}
(r + 2)^{2} a_{r+2,s} &=& -(s + 2)(s + 1)a_{r,s+2} + \sum_{k,l,m,n,p,g}(a_{kl}a^{*}_{mn}
- b_{kl}b^{*}_{mn})\nn\\
&\times&[a_{pg}(p^{2} + g^{2} - 4p - 5g - 2pk - 2gl - 2)\nn\\
&+& a_{p+2,g-2}(p + 2)(p + 2 - 2k) + a_{p-2,g+2}(g + 2)(g + 1 - 2l)]
\end{eqnarray}
and
\begin{eqnarray}
(r + 2)^{2} b_{r+2,s} &=& -(s + 2)(s + 1)b_{r,s+2} + \sum_{k,l,m,n,p,g} (a_{kl}a^{*}_{mn}
- b_{kl}b^{*}_{mn})\nn\\
&\times&[b_{pg}(p^{2} + g^{2} - 4p - 5g - 2pk - 2gl - 2)\nn\\
&+& b_{p+2,g-2}(p + 2)(p + 2 - 2k) + b_{p-2,g+2}(g + 2)(g + 1 - 2l)],
\end{eqnarray}
where $m = r - k - p, 0 \leq k \leq r, 0 \leq p \leq r - k$ with $k$ and $p$ even, and $n = s - l - g$,
$0 \leq l \leq s + 1$, and $-1 \leq g \leq s - l$. These recursive relations could build the whole
power series of $\tilde{\xi}$ and $\tilde{\varsigma}$ from their values on the axis of symmetry
\begin{eqnarray}
\tilde{\xi}(\bar{\rho}=0) = \sum_{i=0}^{\infty} m_{i}\bar{z}^{i},\quad
\tilde{\varsigma}(\bar{\rho}=0) = \sum_{i=0}^{\infty} q_{i}\bar{z}^{i}\label{q0}.
\end{eqnarray}
The values of the multipolar moments of the spacetime are determined in terms of the coefficients
in the series expansion (\ref{q0}). The relations between $a_{ij}$ and $b_{ij}$ can be used in order
to express the moments in terms of $q_{i}\equiv b_{0i}$ and $m_{i}\equiv a_{0i}$.

Following the previous procedure, the multipolar moments for Manko's solution (in agreement with
\cite{mak93}) are $m_{0}=m$, $m_1=i ma$, $m_2=-ma^2$, $m_3= -ima^3$, $q_0=q$, $q_1=i(b+ a q)$,
$q_2=-a(a q + b)$ and $q_3= -ia^2(a q + b)$, whence it follows that parameters $m$, $a$ and $q$
represent, the total mass, the total angular momentum per unit mass and the
total charge, respectively. The parameter $b$ is related to the magnetic dipole $\mu$ by means of
$\mu={\rm{Im}}(q_1)=b+aq$. Then, by defining
\begin{eqnarray}\label{b}
b=aq\left(\frac{g}{2}-1\right),
\end{eqnarray}
with $g$ the Lande factor one can see that the multipolar moment $\mu={\rm{Im}}(q_1)$ reduces to the
classical expression for the magnetic dipole $\mu=\frac{g q J} {2 m}$ \cite{FLS65}. This simple
transformation lets us study particles with an arbitrary $g$-factor, but restricted to the
case of charged particles. An alternative approach to solve this limitation is to consider a more
general solution, {\textit{e.g.}} \cite{MMR95,PRS06}, but due to the cumbersome form of the expressions
this will be study elsewhere. It is important to notice that if we set $b=0$ in (\ref{Ernstaxis}) or
$g=2$ in (\ref{b}), we obtain the Kerr-Newman solution \cite{NC&65}. This fact shows that the asymptotic
expansion of Manko's solution represents the simplest model for studying systems with arbitrary
intrinsic magnetic dipole as atomic nuclei or other subatomic systems.

On the other hand, by performing the asymptotic expansion (\ref{xi-q}) in polar spherical coordinates
with $i\leq 4$ and $j\leq (3-i)$, {\textit{i.e.}} up to the first term including short range interactions,
we get the approximate expressions for the mass and the rotation potentials which are given, respectively,
by the real and imaginary parts of $\xi$
\begin{eqnarray}
\xi&=&\phi_{M}+ i \phi_{J},\label{xi}\\
\phi_{M}/m&=&\left[\frac{1}{r}-\frac{a^{2}}{r^3}P_2\right]+
\left[\frac{q^{2}-m^{2}}{3r^3}(1-P_2)\right]+{\cal{O}}(r^5),\label{phim}\\
\phi_{J}/J&=&\left[\frac{1}{r^2}+\frac{3(a^{2}-m^{2}+q^{2})}{5 r^4}\right]P_1-
\left[\frac{3(a^{2}-m^{2}+q^{2})}{5 r^4}\right]P_3+{\cal{O}}(r^6),\nn
\end{eqnarray}
and the approximate expressions for the electrostatic and the magnetic potentials which are given by the
real and imaginary parts of $\varsigma$, respectively,
\begin{eqnarray}
\varsigma&=&\phi_{E}+ i \phi_{H},\label{varsigma}\\
\phi_{E}/q&=&\left[\frac{1}{r}- \frac{g a^{2}}{2 r^3}P_2\right]+
\left[\frac{q^{2}-m^{2}}{3r^3}(1-P_2)\right]+{\cal{O}}(r^5),\label{phie}\\
\phi_{H}/\mu&=&\left[\frac{1}{r^2}+\frac{3(a^{2}-m^{2}+q^{2})}{5 r^4}\right]P_1-
\left[\frac{3(a^{2}-m^{2}+q^{2})}{5 r^4}\right]P_3+{\cal{O}}(r^6),\nn
\end{eqnarray}
Here, $P_n = P_n(\cos\theta)$ denotes the Legendre polynomial of $n$th degree. It is worth mentioning
that although potentials $\xi$ and $\varsigma$ given in (\ref{xi}) and (\ref{varsigma}) satisfy
(\ref{eq:GenerLaplaceXI}) and (\ref{eq:GenerLaplaceSIGMA}), up to the same order of approximation,
their real parts do not satisfy the Laplace equation in flat space. In the classical theory, static
electric and magnetic fields do not interact, however in Einstein-Maxwell theory they do, and except
in very special cases, there is an interaction tending to introduce a rotation into the spacetime
\cite{bon91,bon06,HG&06}. Expressions above confirm this fact in the post-Newtonian limit. In addition,
it should be pointed out that the effect of a Lande $g$-factor different from $g_{\mathrm{KN}}$ will
be reflected only in the electric potential.

\section{Some consequences of the arbitrary Lande $g$-factor}\label{sec4}
Given that the magnetic dipole for Manko's solution is referred to as $\mu=g q a /2$, let us assume
that the intrinsic spin angular momentum $s$ of any charged lepton, nucleon or nuclei, can be identified
with the angular momentum $J$ of Manko's solution. In the particular case of spin-$1/2$ particles,
the following relation holds
\begin{eqnarray}
\label{equ:aDefQuant}
a = \frac{J}{m}= \frac{\hbar}{2 m}.
\end{eqnarray}
By using the fact that in natural units, e.g., for the electron
$a_{\mathrm{e}} \sim 1.93\times10^{-11} \mathrm{cm} \gg
 q_{\mathrm{e}} \sim 1.38\times10^{-34} \mathrm{cm} \gg
 m_{\mathrm{e}} \sim 6.76\times10^{-56} \mathrm{cm}$,
without lost of generality, the approximate expressions for the gravitational and electrostatic
potentials (\ref{phim},\ref{phie}) can be written as
\begin{eqnarray}\label{equator}
\phi_{M}=\frac{G m}{r}\left[1+\frac{a^2}{c^{2}r^{2}}P_2(\cos\theta)\right],
\quad
\phi_{E}=\frac{k q}{r}\left[1+\frac{g a^2 }{2 c^2 r^{2}}P_2(\cos\theta)\right],
\end{eqnarray}
where $G$ denotes the Newton gravitational constant, $k$ the Coulomb constant and $c$ the
speed of the light. Taking into consideration Eq.~(\ref{equ:aDefQuant}), we can write
$\phi_{M}=\frac{G m}{r}\left[1+\lbar_{\mathrm C}^2P_2(\cos\theta)/4r^{2}\right]$ and
$\phi_{E}=\frac{k q}{r}\left[1+g \lbar_{\mathrm C}^2 P_2(\cos\theta)/8 r^{2}\right]$ to note
that, in this case, corrections are expected at lengths of the order of the reduced Compton
wavelength and additionally the expressions between brackets for the modified potentials become identical
for the case of the Kerr-Newman solution $g_{\rm KN}=2$. The modified
Coulomb potential given above coincides with the potential given by Rosquist (\textit{cf.} Eq.~(14)
in \cite{ros06}), except for the missing $g$ factor and a factor $1/3$.

\begin{figure}[h!]
\begin{center}
\includegraphics[width=10cm]{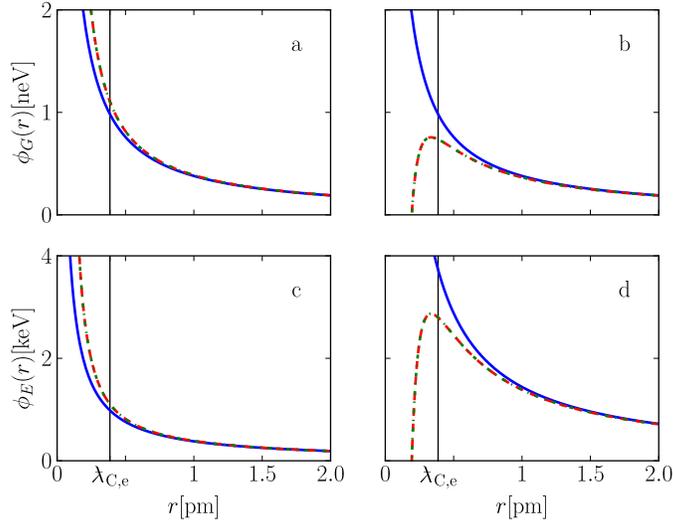}
\caption{Gravitational [(a) and (b)] and electric [(c) and (d)]
potential energy for the classical potentials (continuous line) together with the obtained by using
Kerr-Newman' solution (dot-dashed line, $g_{\mathrm{KN}}=2$) and Manko's solution [dashed line,
$g_{\mathrm{e}}=2.0023193043768(86)$] for the electron. Panels (a) and (c) stand for potentials
on the equator while (b) and (d) do along the spin axis. $\lbar_{\rm C,e}$ denotes the reduced Compton
wavelength of the electron $\lbar_{\rm C,e} = 0.386159$ pm.}
\label{electron}
\end{center}
\end{figure}
\begin{figure}[h!]
\begin{center}
\includegraphics[width=10cm]{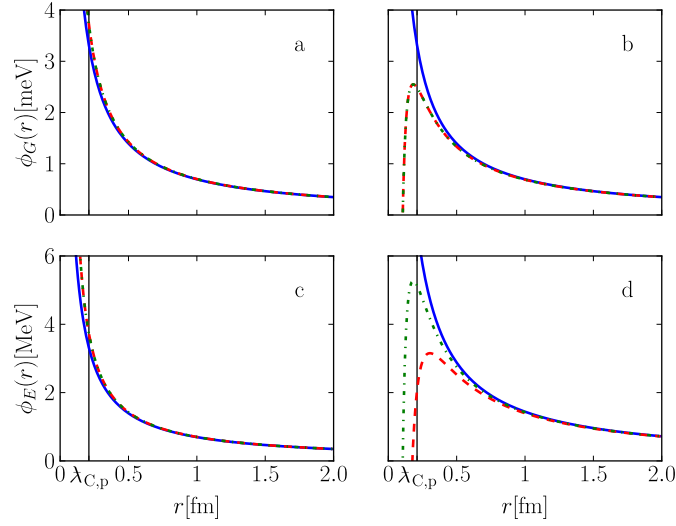}
\caption{Gravitational [(a) and (b)] and electric [(c) and (d)] potential energy for the classical
potentials (continuous line) together with the obtained by using Kerr-Newman' solution (dot-dashed
line, $g_{\mathrm{KN}}=2$) and Manko's solution [dashed line, $g_{\mathrm{p}}=5.585694713(46)$] for
the proton. Panels (a) and (c) stand for potentials on the equator while (b) and (d) do along the
spin axis. $\lbar_{\rm C,p}$ denotes the reduced Compton wavelength of the proton $\lbar_{\rm C,p} =
0.210309$ fm.}
\label{proton}
\end{center}
\end{figure}
In Figs.~(\ref{electron}-\ref{proton}), we plot the classical gravitational and electric potential
energies of a test particle with unit charge and mass together with the corrections predicted from the
Kerr-Newman solution and from Manko's solution. In spite of the fact that the effect of using the real
Lande $g$-factor becomes appreciable only for the electric potential of the proton at length scales bellow
$1$ fm [see Figs.~(\ref{electron}-\ref{proton})], in general, the gravitational and electric interaction
will deviate significantly from its classical form at the reduced Compton scale. As can be seen from
Figs.~(\ref{electron}-\ref{proton}) the differences between the potential shapes along the spin axis
and over the equatorial plane is not an effect introduced by the Lande $g$--factor but rather is a
consequence of the quadrupolar-like symmetry of the Kerr-Newman and Manko solutions. For the gravitational
potential, the quadrupolar mass term does not depend on $g$ and is given by $m_2=-a^2 m$ for both of
the solutions; while for the electric potential, the quadrupolar term is given by
$q_2^{\mathrm{KN}}= - q J^2/m^2$ for Kerr-Newman' solution and by $q_2= - g q J^2/ 2 m^2$ for Manko's
solution, so this will tend to be more marked for the proton.

Since at the Compton-wavelength scale the gravitational field becomes spin dominated rather than mass
dominated, it is expected that some corrections to the hydrogen atom could be experimentally tested
\cite{ros06}. In order to render this argument more quantitative, we calculate the Bohr radius and
the bound state energies of the electron for hydrogen-like atoms. Due to the fact that at distances
of atomic radius the iso-potential surfaces in (\ref{equator}) are practically spherical, we take the
expression over the equator. Under these assumptions, the explicit formulas are given by
\begin{eqnarray}
R&=&\frac{c^2 n^2 \hbar ^2+\sqrt{c^4 n^4 \hbar ^4-3 c^2 \Omega }}{2 k m_{\mathrm{e}} c^2 q^2 Z}
\label{Rb}
\\
E&=&-\frac{2 m_{\mathrm{e}} k^2 c^4 q^4 Z^2 \left(c^2 n^4 \hbar^4 + n^2\hbar^2
\sqrt{c^4 n^4 \hbar ^4-3 c^2 \Omega} - 2\Omega\right)}{\left(c^2 n^2
\hbar ^2 + \sqrt{c^4 n^4 \hbar^4 - 3 c^2 \Omega}\right)^3}
\label{Eb}
\end{eqnarray}
where $\Omega=g_{\mathrm{p}}  a_{\mathrm{p}}^2 k^2 m_{\mathrm{e}}^2 q^4$, $m_{\mathrm{e}}$ is the
electron mass, $g_{\mathrm{p}}$ the Lande $g$-factor for the proton, $Z$ is the atomic number, $n$
the principal quantum number and $a_{\mathrm{p}}$ is the angular momentum per unit mass for the proton.
To get an idea of the order of the possible corrections, here we have assumed that the
total nuclear angular momentum per mass unit is $a_{Z} = \hbar/(2 Z m_{\mathrm{p}})$; however, in
real nuclei this value can certainly be completely different and additionally $m_Z \neq Z m_{\mathrm{p}}$.
Note that after neglecting the spin-dependent terms, {\textit{i.e.}} in the limit
$a_{\mathrm{p}}\rightarrow 0$, expressions (\ref{Rb})-(\ref{Eb}) reduce to the well known classical
formulas. The calculated energy and radius differences between the new expressions
(\ref{Rb})-(\ref{Eb}) and the predicted by the classical Bohr model for the first hydrogen-like
atoms with $n=1$ are listed below in Table \ref{table}.
\begin{table}[h!]
\begin{center}
\begin{tabular}{| c| c| c|}
  \hline \hline
  &$\Delta E $($10^{-10}$ eV) & $\Delta R (10^{-12}${\AA})\\
  \hline\hline
  $\textrm{H}$ & $1.50047$ & $8.75383$ \\
  $\textrm{He}^{+}$ & $6.00188$ & $4.37692$ \\
  $\textrm{Li}^{2+}$ & $13.5043$ & $2.91794$ \\
  $\textrm{Be}^{3+}$ & $24.0075$ & $2.18846$ \\
  $\textrm{B}^{4+}$ & $37.5115$ & $1.75076$ \\
  \hline \hline
\end{tabular}
\vspace{0.5cm}
\caption{Differences between the new expressions (\ref{Rb})-(\ref{Eb}) and the predicted by Bohr
for the energy $\Delta E$ and the radius $\Delta R$ for the first hydrogen-like atoms in the ground
state. Since the measurement uncertainty of the Bohr's radius is $3.6 \times 10^{-10}$\AA
\cite{MTN08} and the uncertainty for the Rydberg constant times $hc$ is $3.4 \times^{-7}$eV
\cite{MTN08}, our corrections for these cases are just behind the uncertainty of the measurements of
the fundamentals constants.}
\label{table}
\end{center}
\end{table}

In spite of, for light atoms these estimations are not promising, based on the tendency, one could
expect that possible corrections to the energy of the ground state of highly ionized heavy atoms
could be tested, e.g., in the case of a molybdenum atom Mo$^{41+}$ the correction to the energy would
be of the order of $0.264685$ $\mu$eV (0.094773 $\mu$eV using $g_{\rm KN}$) and for a gold atom
Au$^{78+}$ the correction of the energy would be of the order of 0.936459 $\mu$eV (0.335291 $\mu$eV
using $g_{\rm KN}$). Although reaching such setup is practically impossible for the gold because it
would imply, e.g., tremendously high temperatures, there already exits data in Literature for H-like
spectra of Mo\footnote{Additional data for H-like spectra of Ti, V, Cr, Mn, Fe, Co, Ni, Cu, and Kr are
also available \cite{SS&00} and some analytic quantum estimations for Rb have been given in \cite{san06}.},
the binding energy for $n=1$ of Mo$^{41+}$ was estimated about 24572.21~$\pm$~0.01~eV using QDE
corrections \cite{SS&00}. Our estimation provides a value of 24000.4405842513 eV while the usual Bohr
expression leaves 24000.440584516 eV, both values differ from the QDE calculated-value, and our
correction is some orders of magnitude below the uncertainty of the value predicted by QDE. However,
it is important to mention that, in this case, the length scales at which this correction takes place,
$R({\mathrm{Mo}}^{41+}) \sim 1.25995$pm, is far from the scale at which deviations are expected,
$a_{\mathrm{Mo}}/c = \hbar/(2 c Z m_{\mathrm{p}}) \sim 8.35 \times 10^{-15}$pm.

Finally, we want to make contact with previous results from the one-loop quantum corrections to the
Newton and Coulomb potential induced by the combination of graviton and photon fluctuations. In
particular, we choose the recent calculations in \cite{BDH03,but06} considering massive charged
spin-$\frac{1}{2}$ fermions. This calculation already contains relativistic corrections, the leading
order corrections are $V_{\rm pN} = G m/c^2 r$ for the relativistic post-Newtonian term and
$V_{\rm q} = G \hbar/ c^3 r^2 = l_{\rm P}^2/r^2$ for the quantum corrections, however, they are by
some order of magnitudes smaller than our leading corrections: $g a^2/ 2 c^2 r^2$ for the Coulomb
potential and $a^2/c^2 r^2$ for the Newtonian potential in (\ref{equator}). At the order of the
reduced Compton wavelength, $V_{\rm pN}$ and $V_{\rm q}$ are of the order of
$10^{-39}$ for the proton and $10^{-45}$ for the electron, while (\ref{equator}) predicts corrections
of the order of $10^{-1}$. This can be appreciated in Figs. (\ref{electron}-\ref{proton}) around the
reduced Compton wavelength of the electron $\lbar_{\rm C,e}$  and of the proton $\lbar_{\rm C,p}$.
Since our corrections come from the rotation of the source, this suggests that angular momentum is
worth considering in quantum gravity calculations in a more detailed way.

\section{Concluding remarks}\label{sec5}
We have studied the influence of the gyromagnetic ratio in the description of the classical fields
of the electron and proton, and showed that the description based on Kerr-Newman solution deviates
significantly for the proton. Based on our assumptions in Sec.~\ref{sec4}, we have obtained that,
although general relativistic effects could be expected in highly ionized heavy atoms, our estimations
could hardly be detected. However, we consider that a more detailed analysis should be carried out, e.g.
taking into account the non-central character of the corrections and considering nuclei with a higher
angular momentum, e.g. in $^{95}_{42}$Mo (spin 5/2), $^{73}_{32}$Ge (spin 9/2) or $^{195\mathrm{m}}_{80}$Hg
(spin 13/2) or explore, e.g., if our corrections would change the hyperfine splitting of hydrogen.
On the other hand, we also point out the necessity of including angular momentum effects in the quantum
description of gravity at the order of at the Compton length scale.
\ack
We thank Professor Bill Bonnor for originally directing our attention to the asymptotic expansion of
the metric element of the rotating charged massive magnetic dipole and subsequent discussions and
comments. Fruitful discussions with L. A. N\'u\~nez and Y. Rodriguez and financial support by DAAD,
Colciencias and Universidad Nacional de Colombia are acknowledged with pleasure. The authors would
like to thank F. Lopez-Suspes for helpful discussions at early stages of this paper.

\bibliographystyle{unsrt}
\bibliography{dlfcqgv2}

\begin{thebibliography}{10}

\bibitem{EG73}
Epstein H and Glaser V.
\newblock {\em Annales Poincare Phys. Theor.}, A19:211, 1973.

\bibitem{HPS00}
Helay\"{e}l-Neto~J A, Penna-Firme~A B, and Shapiro~I L.
\newblock {\em JHEP}, 01:009, 2000.

\bibitem{EL&95}
Elizalde E, Lousto~C O, Odintsov~S D, and Romeo A.
\newblock {\em Phys. Rev. D}, 52:2202, 1995.

\bibitem{gri01}
Grillo N.
\newblock {\em Class. Quantum Grav.}, 18:141, 2001.

\bibitem{bje02}
Bjerrum-Bohr N~E J.
\newblock {\em Phys. Rev. D}, 66:084023, 2002.

\bibitem{BDH03}
Bjerrum-Bohr N~E J, Donoghue~J F, and Holstein~B R.
\newblock {\em Phys. Rev. D}, 67:084033, 2003.

\bibitem{but06}
Butt~M S.
\newblock {\em Phys. Rev. D}, 74:125007, 2006.

\bibitem{fal08}
Faller S.
\newblock {\em Phys. Rev. D}, 77:124039, 2008.

\bibitem{MP68}
Martin~A W and Pritchett~P L.
\newblock {\em J. Math. Phys.}, 9:593, 1968.

\bibitem{isr70}
Israel W.
\newblock {\em Phys. Rev. D}, 2:641, 1970.

\bibitem{NW74}
Newman~E T and Winicour J.
\newblock {\em J. Math. Phys.}, 15:1113, 1974.

\bibitem{bur74}
Burinskii~A Ya.
\newblock {\em Sov. Phys. JETP}, 39:193, 1974.

\bibitem{lop84}
L\'opez~C A.
\newblock {\em Phys. Rev. D}, 30:313, 1984.

\bibitem{MM93}
Mann~R B and Morris~M S.
\newblock {\em Phys. Lett. A}, 181:443, 1993.

\bibitem{bur03}
Burinskii A.
\newblock {\em Phys. Rev. D}, 68:105004, 2003.

\bibitem{AP04}
Arcos~H I and Pereira~J G.
\newblock {\em Gen. Relativ. Gravit.}, 36:2441, 2004.

\bibitem{bur04}
Burinskii A.
\newblock {\em Phys. Rev. D}, 70:086006, 2004.

\bibitem{bur05}
Burinskii A.
\newblock {\em Czech. J. Phys.}, 55:A261, 2005.

\bibitem{ros06}
Rosquist K.
\newblock {\em Class. Quantum Grav.}, 23:3111, 2006.

\bibitem{car68}
Carter B.
\newblock {\em Phys. Rev.}, 174:1559, 1968.

\bibitem{NC&65}
Newman~E T, Couch E, Chinnapared K, Exton A, Prakash A, and Torrence R.
\newblock {\em J. Math. Phys.}, 6:918, 1965.

\bibitem{MTN08}
Mohr~P J, Taylor~B N, and Newell~D B.
\newblock {\em Rev. Mod. Phys.}, 80(2):633, 2008.

\bibitem{mak93}
Manko~V S.
\newblock {\em Phys. Lett. A}, 181:349, 1993.

\bibitem{ern68}
Ernst~F J.
\newblock {\em Phys. Rev.}, 168:1415, 1968.

\bibitem{MS93}
Manko~V S and Sibgatulli~N R.
\newblock {\em Class. Quantum Grav.}, 10:1383, 1993.

\bibitem{sib91}
Sibgatullin~N R.
\newblock {\em Oscillations and waves in strong gravitational and
  electromagnetic fields (Engl. transl.)}.
\newblock Springer, Berlin [orig. Russian, 1984, Nauka, Moscow], 1991.

\bibitem{SA04}
Sotiriou P and Apostolatos A.
\newblock {\em Class. Quantum Grav.}, 21:5727, 2004.

\bibitem{ger70}
Geroch R.
\newblock {\em J. Math. Phys}, 11:2580, 1970.

\bibitem{han74}
Hansen~R O.
\newblock {\em J. Math. Phys}, 15:46, 1974.

\bibitem{PS06}
Pach\'on~L A and Sanabria-G\'omez~J D.
\newblock {\em Class. Quantum Grav.}, 23:3251, 2006.

\bibitem{FLS65}
Feynman~R P, Leighton~R B, and Sands M.
\newblock {\em The Feynman Lectures on Physics, Quantum Mechanics, Vol III}.
\newblock Addison Wesley Publishing Company, Massachusetts, U.S.A., 1965.

\bibitem{MMR95}
Manko~V S, Mart\'in J, and Ru\'iz E.
\newblock {\em Phys. Rev. D}, 36:3063, 1995.

\bibitem{PRS06}
Pach\'on~L A, Rueda~J A, and Sanabria-G\'omez~J D.
\newblock {\em Phys. Rev. D}, 73:104038, 2006.

\bibitem{bon91}
Bonnor~W B.
\newblock {\em Phys. Lett. A}, 158:23, 1991.

\bibitem{bon06}
Bonnor~W B.
\newblock {\em Gen. Relativ. Gravit.}, 38:1063, 2006.

\bibitem{HG&06}
Herrera L, Gonz\'alez~G A, Pach\'on~L A, and Rueda~J A.
\newblock {\em Class. Quantum Grav.}, 23:2395, 2006.

\bibitem{SS&00}
Shirai T, Sugar J, Musgrove A, and Wiese~W L.
\newblock {\em Spectral Data For Highly Ionized Atoms: Ti, V, Cr, Mn, Fe, Co,
  Ni, Cu, Kr, and Mo}.
\newblock J. Phys. Chem. Ref. Data, Monograph No. 8, 2000.

\bibitem{san06}
Sansonetti~J E.
\newblock {\em J. Phys. Chem. Ref. Data}, 35:301, 2006.

\end{thebibliography}

\suppressfloats
\end{document}